\documentstyle[twoside,fleqn,espcrc2]{article}

\input epsf

\newcommand{\DT}{}

\newcommand{\etal}{{\it et al.}}

\newcommand{\LL}{\left\langle}
\newcommand{\RR}{\right\rangle}
\newcommand{\PAR}[2]{ {{\partial{#1}}\over{\partial {#2}}} }

\newcommand{\BE}{\begin{equation}}
\newcommand{\EE}{\end{equation}}
\newcommand{\BEA}{\begin{eqnarray}}
\newcommand{\EEA}{\end{eqnarray}}
\newcommand{\EL}{\nonumber\\}

\newcommand{\pbp}{\bar\psi\psi}
\newcommand{\Plaq}{\Box}
\newcommand{\gbeta}{6/g^2}

\newcommand{\AmS}{{\protect\the\textfont2
  A\kern-.1667em\lower.5ex\hbox{M}\kern-.125emS}}

\title{The $N_t=6$ equation of state for two flavor QCD}

\DT
\DT
\DT
\DT
\DT

\author{ C.~Bernard
\address{{\vskip-0.10in{\hskip 0.07in Department of Physics, Washington
University, St.~Louis, MO 63130, USA \vskip -0.03in}}},
T. Blum
\address{Department of Physics, University of Arizona, Tucson, AZ 85721, USA
\vskip -0.03in},
C.E.~DeTar
\address{Physics Department, University of Utah, Salt Lake City, UT 84112, USA
\vskip -0.03in},
Steven~Gottlieb
\address{Department of Physics, Indiana University, Bloomington, IN 47405, USA
\vskip -0.03in},
U.M.~Heller
\address{SCRI, Florida State University, Tallahassee, FL 32306, USA \vskip
-0.03in},
J.E.~Hetrick$\,\null^{\rm b}$,\\
L.~K\"arkk\"ainen
\address{Nordita, Blegdamsvej 17, DK-2100 Copenhagen \O, Denmark \vskip
-0.03in},
K.~Rummukainen$\,\null^{\rm d}$,
R.L.~Sugar
\address{Department of Physics, University of California, Santa Barbara, CA
93106, USA \vskip -0.03in},
D.~Toussaint$\,\null^{\rm b}$
, and M.~Wingate
\address{Physics Department, University of Colorado, Boulder, CO 80309, USA}
} 

\begin{document}

\begin{abstract}
We improve the calculation of the equation of state for two flavor
QCD by simulating on $N_t=6$ lattices at appropriate values of the couplings
for the deconfinement/chiral symmetry restoration crossover.
For $am_q=0.0125$ the energy density rises
rapidly to approximately 1 ${\rm GeV/fm^3}$ just after
the crossover($m_\pi/m_\rho\approx 0.4$ at this point).
Comparing with our previous result for $N_t=4$~\cite{eos},
we find large finite
$N_t$ corrections as expected from free field theory on finite lattices.
We also provide formulae for extracting the speed of sound from the
measured quantities.
\end{abstract}

\maketitle


\section{INTRODUCTION}

In order to show the existence of the Quark-Gluon Plasma (QGP)
in the aftermath of upcoming heavy-ion collision experiments at RHIC and
CERN or to understand the dynamics of the QGP in the early universe,
one needs as input, among other things, the equation of state for
QCD, {\it i.e.} the energy density and the pressure as a function of
temperature and quark mass.
\DT

\DT
We are continuing our program of computing the equation of state for two
flavor QCD.  Last year we reported results for $N_t=4$~\cite{eos}, and
we are now working at $N_t=6$.  A similar program for the pure gauge
theory is being pursued by the Bielefeld group\cite{BIELEFELD}.
The $N_t=6$
simulations represent a significant increase in computational cost
due to the increased lattice size, smaller quark masses, and corresponding
smaller simulation step sizes.
\DT Step size ($\Delta t$) errors induced by the approximate integration
of the gauge field equations of motion are much more significant at
$N_t=6$ than at $N_t=4$.
They are handled by extrapolation of observables to $\Delta t=0$, and
by running at small step sizes, with a
corresponding large increase in the cost of simulation.

We have surveyed the gauge coupling
and quark mass plane for two flavor QCD in a region relevant to the
nonzero temperature crossover in order to measure the
nonperturbative pressure by integration. The interaction measure
\DT is also calculated and together with the pressure it yields the energy
density.

\section{THEORY}
\DT

\DT A Euclidean $N_s^3\times N_t$ lattice with periodic boundary conditions
has a temperature $T$ and volume $V$
\BEA
V = N_s^3 a^3,\EL
1/T = N_ta,
\label{vandt}
\EEA
where $a$ is the lattice spacing.
Thermodynamic variables are derivatives of the partition function $Z$.
In particular,
the pressure $p$ and energy density $\varepsilon$ are given by
\BE
{\frac{p}{T}} ={\PAR{\log Z}{V}}
\label{pdef}
\EE
and
\BE
\varepsilon V =-{\PAR{\log Z}{(1/T)}}\ .
\label{edef}
\EE

The methods for computing the pressure and integration measure are
discussed in Ref.~\cite{eos}.  The pressure is found by integrating
either the plaquette or $\pbp$.
%
%
\BEA
pa^4 = 2
\int_{\rm cold}^{\gbeta}
[ \LL {\Plaq} \RR
-\LL {\Plaq} \RR_{\rm sym}]d(6/g'^2).
\label{int_plaq}
\EEA
\BEA
pa^4 =
\int_{\rm cold}^{m_qa}
 [\LL \pbp\RR
 - \LL \pbp\RR_{\rm sym}] d(m_q'a).
\label{int_pbp}
\EEA


The interaction measure is given by
\BEA
{\frac{I V}{T}} =
-{\frac{1}{T}}{\PAR{\log Z}{(1/T)}} -3V{\PAR{\log Z}{V}} \EL
=(-a_t {\PAR{}{a_t}} - a_s{\PAR{}{a_s}} )\log Z =
-{\PAR{\log Z}{\log a}}
\label{idef}
\EEA
where $a_s$ and $a_t$ are the spatial and temporal lattice spacings.
The scale dependence in the case of QCD with dynamical quarks leads
to
\BEA
\label{im}
Ia^4 &=& -2\PAR{(\gbeta)}{\log a}
[ \LL \Plaq \RR-  \LL \Plaq \RR_{\rm sym}]\EL
\,&-&{\PAR{(am_q)}{\log a}}
[\LL \pbp \RR - \LL \pbp \RR_{\rm sym}].
\EEA
The derivatives are the usual $\beta$ function and the
anomalous dimension of the quark mass. In the above the subscript ``sym" refers
to symmetric lattices with $N_t=N_s$ which are used for vacuum subtraction.

The knowledge of the non-perturbative pressure and interaction measure
allows us to compute other bulk quantities. The energy
and entropy $s$ become
\BEA
\varepsilon =  I  + 3 p\EL\ \ ,
sT = I  + 4p\ \ .
\label{eands}
\EEA

\section{SOUND SPEED}
Acoustic perturbations travel in the system with a speed $c_s$:
\BE
{\frac{1}{c_s^2}} = {\frac{d\varepsilon}{dp}}.
\label{sounddef}
\EE
One has to take the derivative keeping the physical quark mass
fixed, {\it i.e.} on the line of constant physics. Unfortunately, we
know best only the variations of the energy density and pressure
along lines of constant bare parameters. In order to measure
the correct sound speed one has to use the $\beta$ function to
map the changes in bare parameters to physical changes of temperature.

\DT For QCD with zero chemical potential the only varying
quantity is the temperature $T$.
In the real world the
masses of the quarks do not change and derivatives are taken at
fixed quark mass.
The volume is infinite and it
is divided out of the equations that consider densities.
Hence
\BE
{\frac{1}{c_s^2}} = {\frac{
                            {\frac{d\varepsilon}{dT}}|_{V,m_\pi/m_\rho}
                          }
                          { {\frac{dp}{dT}}|_{V,m_\pi/m_\rho}
                          }
                     }.
\EE

\DT Henceforth we drop the $|_{m_\pi/m_\rho}$ reminder.
With only $T$ varying, the fundamental relation of thermodynamics becomes
\BE
f(T) = \varepsilon(T) - Ts(T) = -p
\EE
or
\BE
\varepsilon(T) = Ts(T) - p(T).
\EE
Then,
\BE
{\frac{d\varepsilon}{dT}} = Ts'(T) + s(T) - p'(T) = Ts'(T),
\EE
where we have utilized a Maxwell relation for the entropy $s$:
\BE
s = {\PAR{p}{T}}|_V = p'(T).
\label{Maxwell}
\EE
In the Maxwell relation we interchanged pressure and
free energy according to
\BE
{\frac{pV}{T}} = \log Z = {\frac{-fV}{T}}.
\label{fandp}
\EE
Hence,
\BE
{\frac{1}{c_s^2}} = {\frac{Ts'(T)}{p'(T)}} = {\frac{d\log(s)}{d\log T}},
\label{logsound}
\EE
where we have used Eq. (\ref{Maxwell}) again.
\DT 

\DT
However, lattice simulations are always done at finite volume, and
the volume varies with the lattice spacing $a$.
To get the correct value one must
take the derivative with respect to temperature with the volume constant.

\DT The derivative with respect to temperature
is problematic. It requires asymmetric
lattice spacing and leads to expressions with asymmetry coefficients.
These in turn, are poorly known in the regime of bare couplings, where
simulations are currently feasible.
\DT It would be advantageous to find an expression that
does not involve the asymmetry coefficients.

Using Eq. (\ref{eands}), one can give the sound speed (\ref{logsound})
with the interaction measure:
\BE
{\frac{1}{c_s^2}} = {\frac{1}{s}}{\frac{dI}{dT}} + 3.
\EE
On the other hand, Eq. (\ref{idef})
allows us to write
\BE
{\frac{1}{c_s^2}}=
3 - {\frac{1}{s}}{\PAR{[{\frac{T }{V}}{\PAR{\log Z}{\log a}}]}{T}}.
\EE

Now, the $T$ derivative is taken at constant $V$. Therefore,
\BE
{\frac{1}{c_s^2}}= 3 -
{\frac{1}{s}}
\left[{
{\frac{1}{V}}{\PAR{\log Z}{\log a}}
      +   {\frac{T}{V}}{\PAR{[{\PAR{\log Z}{\log a}}]}{T}}
}\right]
 .
\label{expandd}
\EE
The last derivative can be given as
\BE
{\PAR{[{\PAR{\log Z}{\log a}}]}{T}} =
{\frac{1}{T}} {\PAR{[{\frac{\varepsilon V}{T}} ]}{\log a} },
\EE
which can be seen by expanding the derivatives with respect to $a$
and $T$ as derivatives with respect to $a_t$ and $a_s$:
\BEA
{\PAR{}{\log a}} &=& a_t {\PAR{}{a_t}} + a_s {\PAR{}{a_s}},
\EL
{\PAR{}{T}} &=& -{\frac{1}{T}} a_t {\PAR{}{a_t}}.
\EEA
We also used the definition
of energy density, Eq. (\ref{edef}),
in the form
\BE
{\frac{\varepsilon V}{T}} = - a_t {\PAR{\log Z}{a_t}}.
\EE
Then Eq.~(\ref{expandd}) becomes
\BE
{\frac{1}{c_s^2}}= 3 +
{\frac{1}{sT}}
\left[{
 I -
   {\frac{T}{V}}
   {\PAR{[{\frac{\varepsilon V}{T}}]}{\log a}}
}\right].
\EE
Using $sT = \varepsilon + p$ and $I = \varepsilon -3p$ this can
be expressed in a rather compact form
\BE
\label{soundformula}
{\frac{1}{c_s^2}}= {\frac{1}{\varepsilon+p}}
\left[{
4\varepsilon -
{\frac{T}{V}} {\PAR{[{\frac{\varepsilon V}{T}}]}{\log a}}
}\right].
\EE
If the system is conformally invariant the energy density
does not depend on the scale and the derivative term
in (\ref{soundformula}) disappears:
\BE
{\frac{1}{c_s^2}}= {\frac{4\varepsilon}{\varepsilon+p}} = 3,
\EE
the relativistic result for a free gas.

For QCD with dynamical quarks an explicit form is
\BEA
{\frac{1}{c_s^2}} =
{\frac{1}{ \varepsilon a^4+ p a^4}}
\left[{
4\varepsilon a^4 -
 ({\PAR{(\gbeta)}{\log a}}) {\PAR{ \varepsilon a^4}{(\gbeta)}}-
}\right.  \EL
\left.{
 ({\PAR{m_qa}{\log a}})      {\PAR{\varepsilon a^4}{(m_qa)}}
}\right]
{}.
\EEA
The derivatives have to be taken on a line of constant physics.

\section{SIMULATIONS}

We have measured $\LL \Plaq \RR$ and $\LL \pbp \RR$
on asymmetric lattices with $N_t=6$ and $N_s=12$, and
on symmetric lattices with $N_t=N_s=12$.

The couplings in the simulations are appropriate for the $N_t=6$
crossover~\cite{CROSS}.
At $am_q=0.0125$ and 0.025 we varied $\gbeta$ between 5.37 and 5.53 and
5.39 and 5.53, respectively. These runs will allow us to extrapolate results
to zero quark mass. At $\gbeta=5.45$ and 5.53, we have varied $am_q$ between
0.01 and 0.1 and 0.0125 and 0.2, respectively. These runs also allow us to
extrapolate to zero quark mass, but in addition they provide a cross check on
the integrations
over $\gbeta$ and give information on the equation of state over a
wide range of quark masses.

Since we are simulating with two flavors of Kogut-Susskind fermions,
the simulations are performed with the refreshed
molecular dynamics R algorithm \cite{Gottlieb87}. For the $N_t=6$ lattices
we ran at least 1800 trajectories after 200 trajectories for thermalization.
On the symmetric lattices we performed at least 800 trajectories after 200
trajectories for thermalization. Each trajectory had unit length in simulation
time.

The R algorithm induces an error
in the observables with a leading term proportional to the square of the step
size, in simulation time, used to integrate the equations of motion
of the gauge fields. In practice, the error is small for each observable.
However, the errors are different on the symmetric and asymmetric lattices,
so they do not cancel from the vacuum subtractions. Moreover, in many
instances the errors are
the same order of magnitude as the vacuum subtracted quantities. Therefore,
they must be eliminated.

\DT In most cases, the step size errors are eliminated by extrapolation.
For each gauge coupling and
quark mass several simulations are run with two or more step sizes. If the step
sizes are small enough, observables depend linearly on the step size
squared. Once this linear dependence is observed, the quantities are
extrapolated
to zero step size. As a rule of thumb, the step size in the R algorithm should
be less than or approximately equal to $am_q$. Roughly speaking, this is
because
in the integration of the gauge momentum, the fermion ``force" to lowest order
is proportional to $1/am_q$. Thus the step taken in simulation time to
update the momentum should be $\simeq am_q$ (or smaller) to keep the change
in the momentum
less than ${\cal O}(1)$. Thus we have used $0.007\leq \Delta t \leq 0.015$ and
$0.015\leq \Delta t \leq 0.03$ for the runs with $am_q=0.0125$ and 0.025,
respectively. For the larger quark mass runs at $\gbeta=5.45$ and 5.53, we
used $0.02 \leq \Delta t \leq 0.03$. Finally, for the smallest quark mass
simulation, $am_q=0.01$ $(\gbeta=5.45)$, we use $\Delta t=0.005$.

\section{EXTRAPOLATIONS}

\DT 

\begin{figure}[htb]
    \vskip -10truept
    \vbox{
     \null\hskip-0.1in\vskip0.7in\epsfxsize=3.0in \epsfbox[0 0 4096
4096]{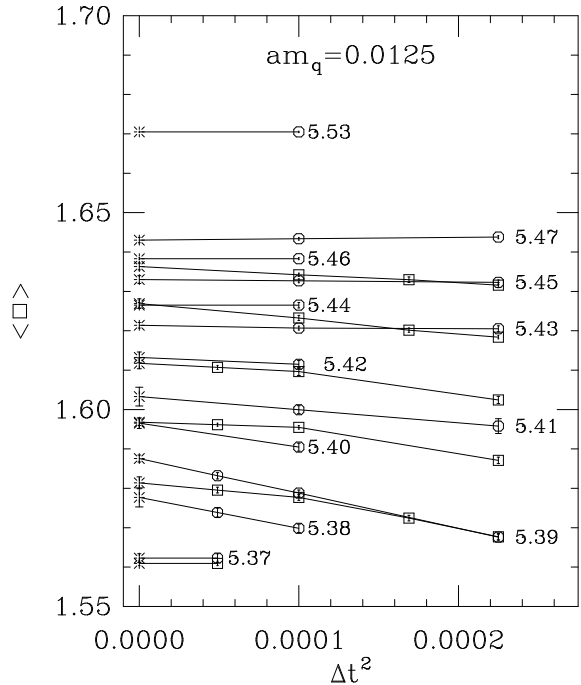}
     \vskip -30truept
     \caption{ The plaquette as a function of $\Delta t^2$ for $am_q=0.0125$.
	Lines are to guide the eye and are not fits. Labels for the
	couplings refer to the hot lattices (octagons). \label{plqm0125}}
    }
\end{figure}

In Fig.~\ref{plqm0125} we show the plaquette dependence on step size
squared for $am_q=0.0125$. Similar results hold for $am_q=0.025$.
Generally, the effects are larger on the symmetric lattices
and at smaller $\gbeta$. For example, at $\gbeta=5.39$, linear behavior sets in
for much smaller step size on the cold lattice than on the hot lattice. In
fact,
for $\Delta t^2\ge 0.0001$, one would conclude that the system is in the
confined
\DT phase; only at smaller step size is a clear separation
visible. From Fig.~\ref{plqm0125},
the following is a reasonable extrapolation procedure. First, on the symmetric
lattices for $5.39\le\gbeta\le5.43$, use only the smallest two step sizes to
extrapolate
to $\Delta t^2=0$. At $\gbeta=5.37$ we effectively assume the system is in the
cold phase and take the values at $\Delta t=0.007$ as the zero step size
extrapolations. For $\gbeta>5.43$ we use all $\Delta t$ values to do the
extrapolations. On the hot lattices we do the following. For each $\gbeta$
where there are multiple step sizes, we use all of them to extrapolate to
$\Delta t^2=0$. For $\gbeta=5.40$ and 5.42 we interpolate the slope from
the neighboring points and use that along with the point at $\Delta t=0.01$ to
extrapolate to zero step size. Note, for $\gbeta\ge5.43$, the slopes are
essentially zero. Therefore, for $\gbeta>5.43$ where only one measurement
is available, we take that value as the zero step size value.

For $\LL\pbp\RR$ the situation is similar to the plaquette with
the following exceptions.
First, after vacuum subtraction, the relative step size errors are not as
significant.
Second, on the hot lattices for $\gbeta>5.39$, we find the slopes with respect
to
$\Delta t^2$ are zero within errors, so we take the zero step size $\LL\pbp\RR$
to be the
value measured at the smallest $\Delta t$ for each of these couplings.

For $am_q=0.025$, the situation is similar, but the smallest step size
runs are still incomplete as of this writing.

\DT Because the cold lattices vary smoothly with $\gbeta$,  we made
cold runs at values of $\gbeta$ separated by $0.02$, while the hot runs were
separated by $\Delta\gbeta=0.01$ near the transition.
Cold observables
at the other couplings were obtained by interpolation.
For $\gbeta\le5.47$, quadratic fits to the zero step size plaquette and $\pbp$
had $\chi^2=2.69$ and 3.18 with three degrees of freedom respectively.
These fits were used for the interpolated values.
To get the symmetric observables
at $\gbeta=5.53$, one can either extrapolate in $\gbeta$ for fixed $am_q$, or
extrapolate in $am_q$ for fixed $\gbeta$. Both give the same result within
errors,
and we simply take the value from extrapolation in $am_q$ as it has a much
smaller
error.
%

\section{PRESSURE}

The results for $\langle\pbp\rangle$ and $\langle\Plaq\rangle$ from the
previous section can now be
integrated to yield the pressure.

\begin{figure}[htb]
    \vskip -10truept
    \vbox{
    \null\hskip-0.1in\epsfxsize=3.0in \epsfbox[0 0 4096 4096]{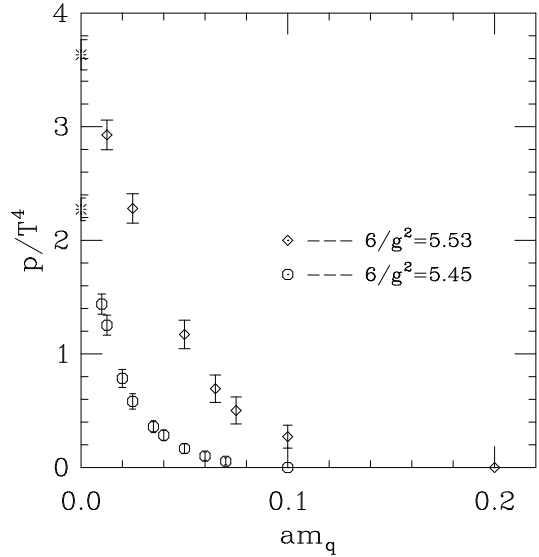}
     \vskip -30truept
     \caption{ The pressure from integration of ${<\pbp>}$ with respect to
	$am_q$. The bursts are extrapolations to zero quark mass. \label{pvsam_q}}
    }
\end{figure}

To begin consider $\langle\pbp\rangle$ as a function of $am_q$. Using
Eq.~(\ref{int_pbp}),
we find the pressure as a function of $am_q$ (at $\gbeta=5.45$ and 5.53).
The result is shown in Fig.~\ref{pvsam_q}. We also want the pressure at
$am_q=0$ which is found by setting $\pbp(0)=0$ and continuing the integration
to $am_q=0$. At $\gbeta=5.45$, a linear fit to the data for $am_q\le0.025$
that is constrained to go through the origin has $\chi^2=4.9$ for three
degrees of freedom. For $\gbeta=5.53$, we only have measurements at
$am_q=0.0125$,
0.025, and 0.05 for which the linear fit does not work well. However, a
quadratic
fit constrained to go through the origin has $\chi^2=0.49$ for one degree of
freedom.
The pressure at $am_q=0$ calculated in this manner is also shown in
Fig.~\ref{pvsam_q}.

\begin{figure}[htb]
    \vskip -10truept
    \vbox{
    \null\hskip-0.1in\epsfxsize=3.0in \epsfbox[0 0 4096 4096]{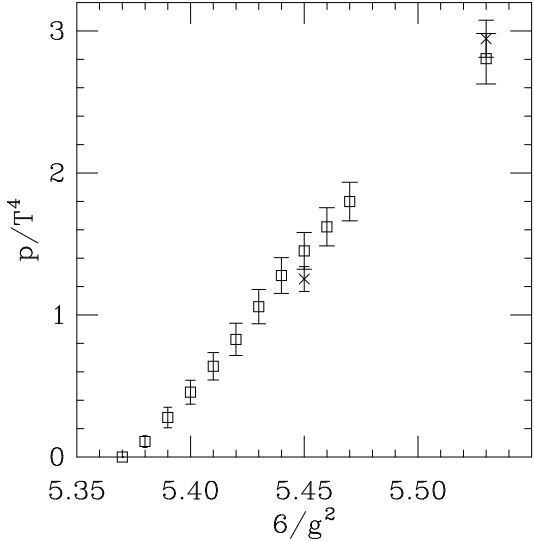}
     \vskip -30truept
     \caption{ The pressure from integration of $\langle\Plaq\rangle$ with
respect
	to $\gbeta$ ($am_q=0.0125$). The crosses are from the $\langle\pbp\rangle$
integration.
     \label{pvsbeta}}
    }
\end{figure}

Next we integrate the plaquette with respect to $\gbeta$ to obtain the pressure
as a function of $\gbeta$ at fixed $am_q$. The result for $am_q=0.0125$ is
shown
in Fig.~\ref{pvsbeta}. The pressure rises smoothly through the crossover region
as it
must. The results from the quark mass integrations are also
shown in Fig.~\ref{pvsbeta}, and are in agreement with the $\gbeta$
integration. This is a good check
on our analysis, since for the most part the two integrations are independent.
In particular, the integration with respect to $\gbeta$ is sensitive to the
step
size extrapolations while the $am_q$ integration is not.

\section{INTERACTION MEASURE}

\begin{figure}[htb]
    \vskip -10truept
    \vbox{
    \null\hskip-0.1in\epsfxsize=3.0in \epsfbox[0 0 4096 4096]{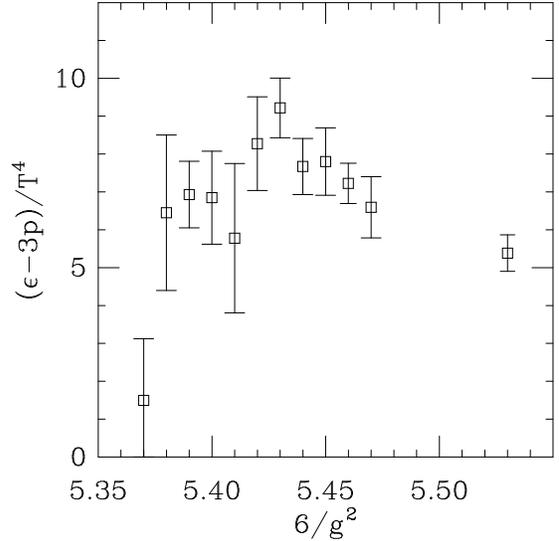}
     \vskip -30truept
     \caption{ The interaction measure for {$am_q=0.0125$}.
     \label{ivsbeta}}
    }
\end{figure}

The interaction measure is given by Eq.~(\ref{im}). For
the $\beta$ function we use our previous result calculated from the $\pi$ and
$\rho$
masses at various values of $\gbeta$ and $am_q$~\cite{eos}.
The result is shown in Fig.~\ref{ivsbeta}. It rises
sharply from zero in the cold phase to some maximum and then begins to drop
off.
For a noninteracting plasma, $I$ is zero since $\varepsilon=3 p$ which is
certainly not the case at the largest value of $\gbeta$ in our
simulations. From asymptotic freedom we expect the system to approach
a noninteracting plasma at very high temperature.

\begin{figure}[htb]
    \vskip -10truept
    \vbox{
    \null\hskip-0.1in\epsfxsize=3.0in \epsfbox[0 0 4096 4096]{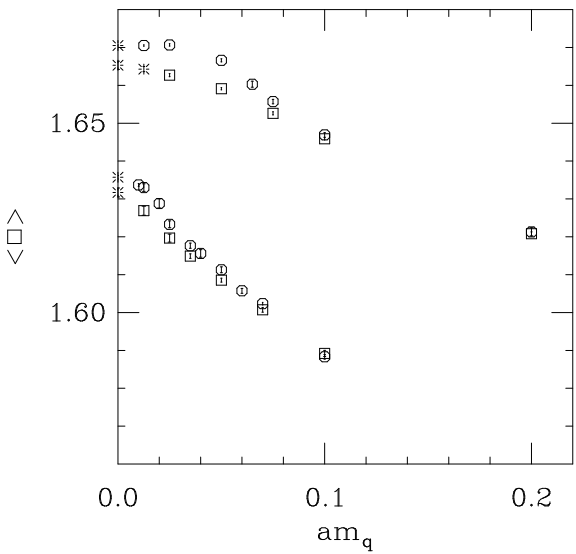}
     \vskip -30truept
     \caption{ The plaquette as a function of the quark mass.
	The lower(upper) curves are for $\gbeta=5.45(5.53)$. The bursts
	are results from fits to the data except for the hot curve
	(octagons) at $\gbeta=5.53$ where the measured value at
	{$am_q=0.0125$} is taken as the zero quark mass result.
        \label{plaqvsam_q}}
    }
\end{figure}

\DT The interaction measure at $am_q=0$ depends only on the plaquette since the
anomalous dimension of the quark mass is zero at $am_q=0$. To extrapolate
the plaquette to zero quark mass we use the fact that its slope
is just its correlation with $\pbp$,
\BEA
\PAR{\LL\Plaq\RR}{(am_q)}=\LL\Plaq \pbp\RR - \LL\Plaq\RR\LL\pbp\RR.
\EEA
Since $\pbp$ is discontinuous at the origin in the broken phase and
continuous in the chirally symmetric phase, we expect a cusp at the
origin for the cold lattices
\DT and zero slope for the hot lattices\cite{eos}.
At $\gbeta=5.45$ a linear fit on the
cold lattices for $am_q\le0.05$ gives $\chi^2=1.6$ with two degrees of freedom.
On the hot lattice a quadratic fit constrained to zero slope at the origin
gives $\chi^2=0.74$ with two degrees of freedom. At $\gbeta=5.53$ the
situation is
less satisfactory. For the cold lattices, a quadratic fit to the data
with $am_q\le0.1$ has $\chi^2=0.09$. Note the smallest quark mass here is
0.025, and we also use this fit for the cold observables at $am_q=0.0125$.
On the hot lattices the data seem to reach a maximum at nonzero quark mass,
so we take the measurement at the smallest quark mass to be the extrapolated
value. Similar
behavior at $\gbeta=5.53$ was observed in our earlier $N_t=4$ study, which may
indicate a systematic error. Naively one expects finite volume effects to order
the lattice which is opposite to the observed behavior. The data and the fit
results are shown in Fig.~\ref{plaqvsam_q}.

\section{EQUATION OF STATE}

\begin{figure}[htb]
    \vskip -10truept
    \vbox{
    \null\hskip-0.1in\epsfxsize=3.0in \epsfbox[0 0 4096 4096]{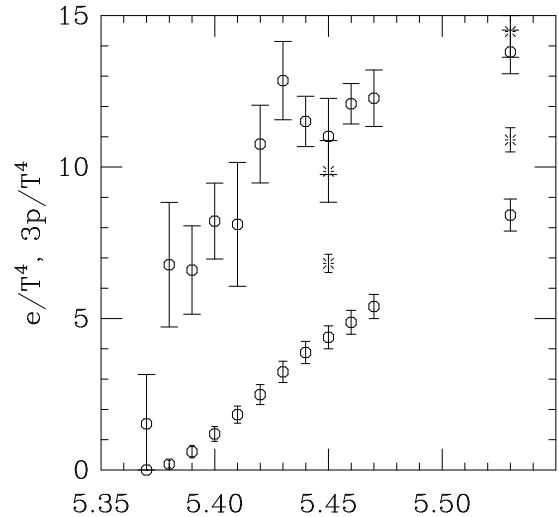}
     \vskip -30truept
     \caption{The ${N_t=6}$ equation of state for ${am_q=0.0125}$ and
	extrapolations to {$am_q=0$}(bursts). The upper curve is $\varepsilon/T^4$.
     \label{eosm0125}}
    }
\end{figure}

In Fig.~\ref{eosm0125} we show the $N_t=6$ equation of state as a function
of $\gbeta$ for $am_q=0.0125$. The zero quark mass extrapolations are also
shown.
At the largest value of $\gbeta$, $\varepsilon$ is still much larger than $3p$;
for $am_q=0$, $3p$ is approximately $75\%$ of $\varepsilon$. On the other hand,
after a rapid rise, $\varepsilon/T^4$ more or less levels off for
$\gbeta\ge5.43$.
This corresponds to an energy density of about 1 ${\rm GeV/fm^3}$ at
$\gbeta=5.43$
(we use the $\rho$ mass to convert $\gbeta$ to temperature). $m_\pi/m_\rho$
at this point is slightly less than 0.4.

\begin{figure}[htb]
    \vskip -10truept
    \vbox{
    \null\hskip-0.3in\epsfxsize=3.2in \epsfbox[0 0 4096 4096]{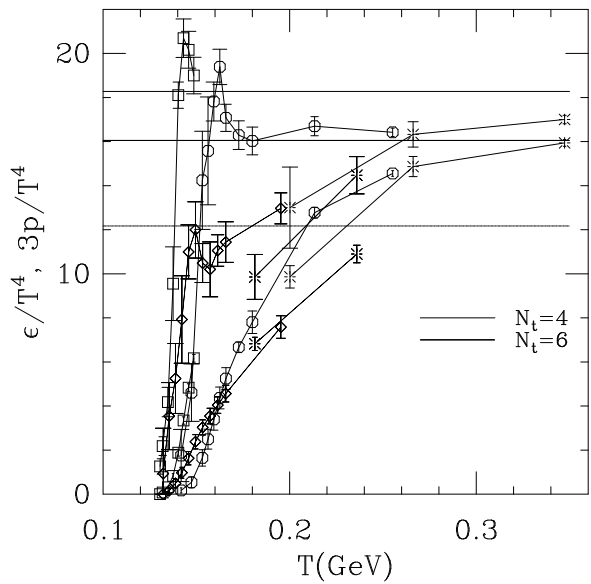}
     \vskip -30truept
     \caption{ Comparison of the equation of state for $N_t=4$ (solid lines)
	and 6 (dashed lines).
	The results shown are for $am_q=0.0125$
	(diamonds), 0.025 (octagons), and 0.1 (squares).
	Bursts are extrapolations to $am_q=0$.
	The horizontal lines give the Stefan-Boltzmann law for $N_t=4$, 6,
	and the continuum (lowest line).
        \label{eos}}
    }
\end{figure}

In Fig.~\ref{eos} we compare the $N_t=4$ and 6 equations of state. There
is a large finite size effect as expected from the free field results on
finite lattices (also shown). For example, $p/T^4$ for
$am_q=0$ differs by about $15\%$. There appears to be a sizable quark mass
effect
as well (the $N_t=4$ results are for $am_q=0.1$ and $0.025$). For $N_t=6$
the approach to the Stefan-Boltzmann law is unclear since we do not have
data at very high temperatures.
The prominent peak in $\varepsilon/T^4$ for $N_t=4$ just after the crossover
\DT
is much smaller at $N_t=6$.
We have plotted the
results versus temperature by using the $\rho$ mass to set the scale. The
crossover temperature is around 150 MeV and is rather insensitive
to $N_t$ and $am_q$.
\DT In a recent preprint, Asakawa and Hatsuda have pointed out that many
features of this equation of state are constrained by fundamental
thermodynamic relations\cite{ASAKAWA}.

\section{SOUND SPEED}

\begin{figure}[htb]
    \vskip -10truept
    \vbox{
    \null\hskip-0.1in\epsfxsize=3.0in \epsfbox[0 0 4096 4096]{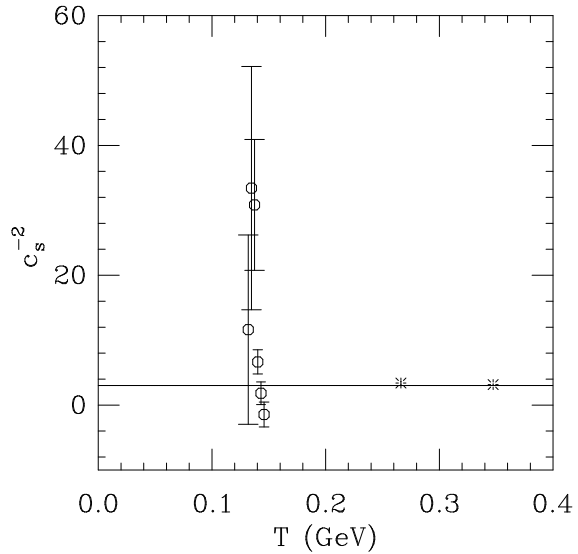}
     \vskip -30truept
     \caption{ The inverse sound speed squared for the $N_t=4$ system at
               $m_qa = 0.1$. \label{soundfig} }
    }
\end{figure}

The sound speed is a quantity that depends on the second derivative of
the partition function. Therefore it is more difficult to get its value
than the values for the thermodynamic variables discussed so far.
Even if our formulation can avoid the asymmetry coefficients there is
an awkward derivative of the energy density in the formula.

To measure the change in the mass as accurately as possible we
performed a set of simulations at $am_q= 0.09$ with $N_t=4$ lattices
in addition to our old data at  $am_q= 0.1$. The result is shown
in Fig.~\ref{soundfig}. Close to the transition the error in
the derivative of the energy density overwhelms our data and we
have large error bars. For large temperatures,
we see that the speed approaches the ideal gas value.

\section{CONCLUSION}

We have calculated the equation of state for QCD with two flavors of quarks
on $N_t=6$ lattices. The algorithm used to generate gauge configurations
introduces a step size error in observables that must be removed by
extrapolation. The added computational cost is significant.
For $am_q=0.0125$, we find the energy density just after the crossover
to be roughly 1 ${\rm GeV/fm^3}$, and for $T$ almost twice the critical
\DT
value, the highest that we simulated, three times the pressure is
only $60\%$ of the energy density.
\DT
We find large effects of nonzero lattice spacing from comparison with
results at $N_t=4$, as expected from free field theory.

\DT
After the completion of runs at $am_q=0.025$, we will complete our
extrapolation to zero quark mass.  It still remains to eliminate
remaining lattice size effects, include the effects of the strange
quark, eliminate any effects of finite volume, and include the effect of
nonzero net quark density.

This work was supported by the US DOE and NSF.  Computations were done
at the San Diego Supercomputer Center, the Cornell Theory Center, and Indiana
University.

\end{document}